\documentclass[journal]{IEEEtran}
\usepackage{amsmath,amsfonts}
\usepackage{algorithmic}
\usepackage{algorithm}
\usepackage{array}
\usepackage{amssymb}
\usepackage[caption=false,font=normalsize,labelfont=sf,textfont=sf]{subfig}
\usepackage{booktabs}
\usepackage{tabularx}
\usepackage{textcomp}
\usepackage{stfloats}
\usepackage{url}
\usepackage{verbatim}
\usepackage{graphicx}
\usepackage{cite}
\usepackage{orcidlink}
\hypersetup{hidelinks} 
\usepackage{xcolor}

\begin{document}

\title{Attention-Enhanced Prompt Decision Transformers for UAV-Assisted Communications with AoI}


\author{Chi~Lu, Yiyang~Ni, Zhe~Wang, Xiaoli~Shi, Jun~Li, Shi~Jin
\thanks{Chi Lu is with the School of Electronic and Optical Engineering, Nanjing University of Science and Technology, Nanjing 210094, China.	(e-mail: luchi2023@njust.edu.cn).
Yiyang Ni is with the Jiangsu Second Normal University, Nanjing 210013, China (e-mail: niyy@jssnu.edu.cn). 
Jun Li and Shi Jin are with the National Mobile Communications Research Laboratory, Southeast University, Nanjing 210096, China (e-mail: jun.li@seu.edu.cn; jinshi@seu.edu.cn).
Zhe Wang is with the School of Computer Science and Engineering, Nanjing University of Science and Technology, Nanjing 210094, China (e-mail: zwang@njust.edu.cn). 
Xiaoli Shi is with the China Aerospace Science and Industry Corporation Intelligent Technology Research Institute, Beijing 100854, China (e-mail: shixiaoli20220003@126.com).
}}


\maketitle

\begin{abstract}
Decision Transformer (DT) has recently demonstrated strong generalizability in dynamic resource allocation within unmanned aerial vehicle (UAV) networks, compared to conventional deep reinforcement learning (DRL). However, its performance is hindered due to zero-padding for varying state dimensions, inability to manage long-term energy constraint, and challenges in acquiring expert samples for few-shot fine-tuning in new scenarios. To overcome these limitations, we propose an attention-enhanced prompt Decision Transformer (APDT) framework to optimize trajectory planning and user scheduling, aiming to minimize the average age of information (AoI) under long-term energy constraint in UAV-assisted Internet of Things (IoT) networks. Specifically, we enhance the convenional DT framework by incorporating an attention mechanism to accommodate varying numbers of terrestrial users, introducing a prompt mechanism based on short trajectory demonstrations for rapid adaptation to new scenarios, and designing a token-assisted method to address the UAV's long-term energy constraint. The APDT framework is  first pre-trained on offline datasets and then efficiently generalized to new scenarios. Simulations demonstrate that APDT achieves twice faster in terms of convergence rate and reduces average AoI by $8\%$ compared to conventional DT.

\end{abstract}


\section{Introduction}
\IEEEPARstart{W}{ith} the rapid advancement of Internet of Things (IoT) technologies, volume of available data has surged, facilitating the widespread applications of AI algorithms in areas such as intelligent transportation, environmental monitoring, and urban management~\cite{weiPersonalizedFederatedLearning2023, maTrustedAIMultiagent2023}. In these scenarios, real-time communications is crucial for various AI applications. Hence, the age of information (AoI) is defined as an important metric for evaluating the efficiency of communications~\cite{yatesAgeInformationIntroduction2021}, which measures the time interval between information generation and reception. By minimizing the long-term AoI, the freshness of the collected data can be guaranteed. 

As a useful approach for communications, unmanned aerial vehicles (UAVs) have gained popularity~\cite{yinDecentralizedFederatedReinforcement2023, huAoIMinimalTrajectoryPlanning2021}, owing to their flexibility and line-of-sight transmission advantages. In an UAV-assisted communications system, path planning and user scheduling are usually optimized for minimizing the AoI~\cite{huAoIMinimalTrajectoryPlanning2021}. However, most methods are based on convex optimizations, necessitating a priori global information of the environment, such as channel states and user locations. This kind of information, however, is difficult to obtain in fast changing environments. Deep reinforcement learning (DRL), as a stochastic optimization technique, is capable of adaptively learning optimal strategies through trial and error. In~\cite{liuAverageAoIMinimization2022}, a deep Q-network (DQN)-based path planning algorithm to minimize the average AoI for ground nodes is proposed. In~\cite{samirAgeInformationAware2020}, Deep Deterministic Policy Gradient (DDPG) algorithm is utilized to optimize UAV trajectory and scheduling in vehicular networks with dynamic vehicle arrivals and departures, aiming to minimize the expected weighted sum AoI. 

Howerver, when the scenario changes (e.g., user density shifts), these DRL models typically need to be retrained from scratch, which is both computationally expensive and time-consuming. Decision Transformer (DT)~\cite{zhangDecisionTransformersWireless2025} utilizes the Transformer's capabilities in comprehension and generalization by reformulating decision-making as a sequence prediction task, learning generalized minimize AoI decision models through autoregressive prediction. Nonetheless, DT typically uses zero-padding to handle changing state dimensions~\cite{zhangDecisionTransformersWireless2025} (e.g., varying numbers of users), leading to suboptimal AoI minimization in complex new scenarios. Additionally, DT is unable to handle long-term energy constraint, and it is challenging to obtain expert samples for few-shot fine-tuning in new scenarios.

To address the above issues discussed, we propose an attention-enhanced prompt Decision Transformer (APDT) framework. The main contributions are summarized as follows.
\begin{itemize}
    \item We introduce an attention mechanism for dynamically aggregating the state spaces from the UAV and users, effectively handling the varying state space dimensions caused by changes in the number of terrestrial users. This ensures that the model can prioritize users based on their relevance, improving the AoI minimize decision.
    
    \item We introduce a prompt mechanism that inserts demonstration sequences with scenario information into the trajectory’s initial part, enabling rapid adaptation to new scenarios. Additionally, we propose a token-assisted method that incorporates energy tokens, which learn the mapping from long-term energy consumption to actions, ensuring the satisfaction of long-term energy constraint.
    \item The proposed APDT framework achieves $4.5$ times faster convergence and reduces the average AoI by over $15\%$ compared to conventional DRL, and demonstrates over $8\%$ lower average AoI with twice faster convergence compared to conventional DT.
\end{itemize}

\section{System Model}
\begin{figure}[!t]
	\centering
	\includegraphics[width=2.6in]{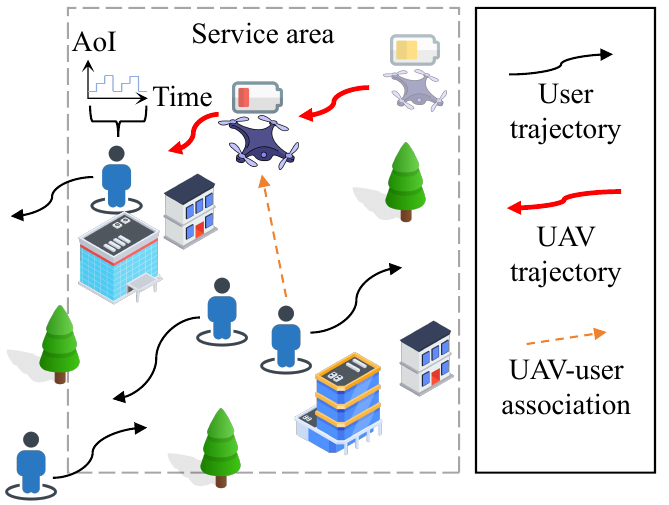}
	\caption{UAV-assisted communications scenario.}
	\label{fig_1}
\end{figure}

In this letter, we consider uplink communication from terrestrial mobile users assisted by an UAV, as illustrated in \figurename~\ref{fig_1}. The duration of UAV's task execution is divided into $T$ evenly spaced time slots, with each time slot $t\in\{1,2,\dots,T\}$ lasting for a duration of $\delta$. Let $\mathcal{U} = \left\{1, 2, \dots, U\right\}$ represent the set of all users, where the horizontal position of each user at time slot $t$ is denoted by $\boldsymbol{q}_u(t), \forall u \in \mathcal{U}$. 

The UAV provides uplink communication services within its designated service area. Unlike prior work with a fixed user set, we allow the user numbers to vary as users enter or leave the service area. Let $\mathcal{K}(t) = \left\{1, 2, \dots, K(t)\right\}$ represent the set of users within the UAV's service area at time slot $t$, where $K(t)$ is the number of users in the area at that time slot. To measure the population density within the service area, we define average user density $\rho$ as the average number of users within the service area over $T$ time slots, calculated as $\rho=(1/T)\sum_{t=1}^TK(t)$. At each time slot $t$, the UAV flies to the target location at an altitude of $H$, then hovers, establishes connections with the users in its service area, and performs uplink communication tasks. The UAV’s horizontal coordinates at time slot $t$ are denoted as $\boldsymbol{q}(t)=(x(t), y(t)) \in \mathbb{R}^2$.

\subsection{UAV Energy Model}
The UAV's movement is divided into two phases, i.e., flying and hovering. First, at time slot $t$, the UAV selects a target user within the service area, represented by $\xi(t) = u $, and determines the destination based on the flight distance $d(t)$ and flight angle $\phi(t)$. The UAV moves to this destination at a maximum speed of $V^{\max}$, with the flight time given by $\delta^{\text{f}}(t)=\frac{d(t)}{V^{\max}}$, and the UAV’s next position \( \boldsymbol{q}(t+1) \) is updated based on its current position and the calculated by $\boldsymbol{q}(t+1) = \boldsymbol{q}(t) + \left(d(t) \cos \phi(t), d(t) \sin \phi(t)\right)$. Once the destination is reached, the UAV hovers to perform uplink communication services for the selected user, with a service duration of $\delta^{\text{h}}(t)=\delta-\delta^{\text{f}}(t)$. The power for the flying and hovering phases are $P^{\text{f}}$ and $P^{\text{h}}$, respectively~\cite{liuPathPlanningUAVMounted2020}. Therefore, the total energy consumption of the UAV at time slot $t$ is $E(t)=P^\text{f}\delta^\text{f}(t)+P^\text{h}\delta^\text{h}(t)$.

\subsection{Age of Information Model}
The AoI of a user is defined as the time elapsed since the user’s most recent data packet, collected by the UAV, was generated. Without loss of generality, the initial AoI of user $u \in \mathcal{K}(t)$ is set as $A_u(0) = 1$. At each time slot \( t \), if user \( u \) is selected for uplink communication by the UAV, a new data packet is generated at the beginning of time slot \( t \) with a size denoted by \( N_u(t) \), and the AoI is updated using
\begin{equation}
A_u(t+1)=\left\{\begin{array}{ll}
A_u(t)+1, & \mathbb{I}_u(t)=0 \\
1, & \mathbb{I}_u(t)=1.
\end{array}\right.
\end{equation}
where \( \mathbb{I}_u(t) \) is a indicator function indicating the service status of user \( u \). Specifically, \( \mathbb{I}_u(t) = 1 \) means the user is successfully served, and \( \mathbb{I}_u(t) = 0 \) indicates the user is not successfully served.  We focus on the average AoI of users as the primary performance metric in this letter,  where at time slot $t$, the average AoI is defined as
\begin{equation}
\bar{A}(t) = \frac{\sum_{u \in \mathcal{K}(t)} A_u(t)}{K(t)}.
\end{equation}
When a new user \( u \) arrives at time slot \( t \), we include \( u \) in \( \mathcal{K}(t) \) and initialize its AoI as \( A_u(t) = \bar{A}(t)\), where \( \bar{A}(t) \) is the average AoI of the current users in the UAV's service area. This initialization ensures that new users are assigned a comparable priority to existing users, avoiding prolonged periods of neglect caused by their lower initial AoI. Once a user leaves, it is removed from \( \mathcal{K}(t+1) \).

Since the UAV can only transmit data to the user during the hovering time \( \delta^{\text{h}} \), we define a required transmission rate \( \tilde{R}_u(t) = \frac{N_u(t)}{\delta^{\text{h}}} \). User \( u \) is considered successfully served ($\mathbb{I}_u(t)=1$) if the actual channel can support this rate, i.e., \( \tilde{R}_u(t) \leq R_u(t) \), where \( R_u(t) \) is the achievable rate, given by
\begin{equation}
    R_u(t)=B \log _2\left(1+\frac{h_u(t)P_u}{\sigma^2}\right),
\end{equation}
where $B$ represents the channel bandwidth, $P_u$ is the transmission power of user $u$, $\sigma^2$ denotes the noise power, and $h_u(t)$ is the channel power gain between the UAV and user $u$. Obviously, AoI is a function of the UAV's hovering duration and user selection.

\section{Problem Formulation}
Our objective is to minimize the long-term average AoI of users by jointly optimizing the UAV's trajectory $\boldsymbol{Q} = \{\boldsymbol{q}(1), \boldsymbol{q}(2), \dots, \boldsymbol{q}(T)\}$ and the user selection $\boldsymbol{\Xi} = \{\xi(1), \xi(2), \dots, \xi(T)\}$ subject to the UAV's energy constraint. The optimization problem can be formulated as
\allowdisplaybreaks
\begin{subequations}
\begin{align}
\text { P1: } & \min _{\boldsymbol{Q}, \boldsymbol{\Xi}} \frac{1}{T} \sum_{t=1}^{T} \bar{A}(t),
\\
\text { s.t. } & \mathcal{C} 1: \left|x(t)\right| \leq x^{\max}, \quad \forall t
\\
& \mathcal{C} 2: \left|y(t)\right| \leq y^{\max}, \quad \forall t
\\
& \mathcal{C} 3: 0 \leq d(t) \leq d^{\max}, \quad \forall t
\\
& \mathcal{C} 4: 0 \leq \phi(t) < 2\pi, \quad \forall t
\\
& \mathcal{C} 5: \sum_{t=1}^T{E(t)} < E^{\text{max}}.
\end{align}
\end{subequations}
Here, $x^{\text{max}} $ and $y^{\text{max}}$ represent the boundaries of the UAV’s service area along the x-axis and y-axis, respectively. $d^{\text{max}}$ represents the maximum flight distance of UAV, and $E^{\text{max}}$ represents the UAV's  energy threshold.

To address the challenges of dynamically adding and removing users without incurring the drawbacks of traditional approaches (e.g., dynamic sliding window), which often require prior knowledge of the environment model, incur high computational cost, and can lead to suboptimal solutions due to frequent reinitialization, we reformulate the optimization problem $\text{P1}$ as a constrained Markov decision process (CMDP) and solve it using DT framework.

We represent a CMDP as a five-tuple \((\mathcal{S}, \mathcal{A}, \mathcal{R}, \mathcal{C}, \mathcal{T})\). Here, \(\mathcal{S}\) represents the state space, containing all possible states \(s(t) \in \mathcal{S}\) at any time \(t\). \(\mathcal{A}\) represents the action space, containing all feasible actions \(a(t)\in \mathcal{A}\) at any time \(t\). \(\mathcal{R}\) and \(\mathcal{C}\) represent the reward and cost functions, describing the reward \(r(t)\) and cost \(c(t)\) obtained when executing action \(a(t)\) in state \(s(t)\). \(\mathcal{T}\) represents the state transition probability function that maps the current state-action pair $(s(t),a(t))$ to the next state $s(t+1)$. Under the CMDP framework, we define the states, actions, rewards, and costs as follows:

\begin{itemize}
    \item \textbf{State}: The state at time \( t \) includes the UAV's position, the users' positions, and the users' AoI, represented as $s(t) = \{q(t), \{q_u(t), A_u(t)\}_{u \in \mathcal{K}(t)}\}$.
    \item \textbf{Action}: The UAV agent's action at time slot \( t \) consists of the UAV's flight distance, flight angle, and the UAV-user association, represented as  $a(t) = \{d(t), \phi(t), \xi(t)\}$. Linear method is employed to map the discrete action $\xi(t)$ into a continuous value, allowing for uniform processing alongside the continuous actions $d(t)$ and $\phi(t)$ to achieve unified handling.
    \item \textbf{Reward}: To align the DRL objective of maximizing expected cumulative r   eturn with the AoI minimization objective, we define the reward at time \( t \) as the negative of the average AoI, i.e., $r(t) = -\bar{A}(t)$.
    \item \textbf{Cost}: The system's cost at time \( t \) is defined as the UAV's energy consumption, i.e.,  $c(t) = E(t)$.

\end{itemize}

After reformulating the optimization problem P1 as a CMDP problem, and considering the lack of generalization ability and difficulty in adapting to new scenarios of conventional DRL methods, we adopt the DT architecture. DT treats DRL as a sequence modeling task, where the trajectories of states, actions, rewards, and costs are modeled as sequential data. By leveraging the strong generalizability of the transformer architecture and employing a supervised learning approach, DT captures long-range dependencies, avoids iterative Bellman updates, and enables policy transfer to new scenarios with few-shot learning, all without the need for frequent retraining. Specifically, each trajectory is represented as a sequence of states $s(t)$, actions $a(t)$, reward-to-go $R(t) = \sum_{t'=t}^T r_{t'}$, structured as $\{R(1), s(1), a(1), \dots, R(T), s(T), a(T)\}$. During deployment, DT generates actions based on the expected cumulative reward set for the scenario.
\section{Proposed Attention-enhanced Prompt Decision Transformer (APDT) Generalization Policy}

\begin{figure}[!t]
	\centering
	\includegraphics[width=3.5in]{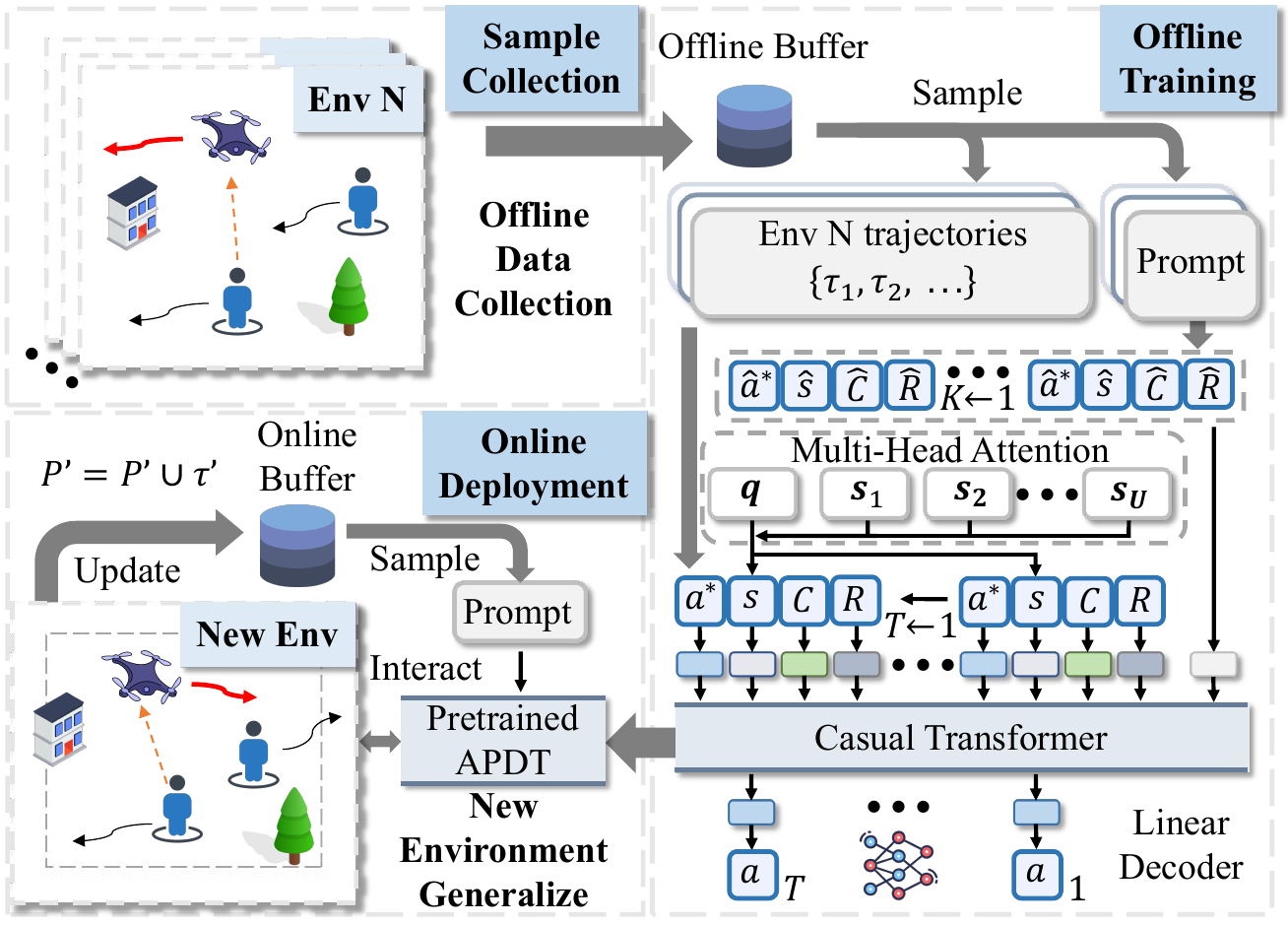}
	\caption{An overview of the APDT framework. Initially, an offline dataset is collected from scenarios with fixed user populations, generating expert trajectories for offline pretraining. Subsequently, the APDT model utilizes attention mechanism and prompt technique to facilitate rapid generalization to new scenarios with varying user populations. Additionally, a token-assisted approach is employed to satisfy long-term energy constraint.}
	\label{fig_2}
\end{figure}

Although DT outperforms classic DRL, it cannot enforce long‑term energy constraints. Few‑shot fine‑tuning also suffers because expert data in a new scenario are scarce.  Finally, conventional DT employs zero-padding to manage state dimensionality changes caused by dynamic user arrivals and departures. However, this approach can lead to suboptimal decisions, as processing uninformative padding distorts the state representation, thereby hindering accurate assessment of the relative AoI and positions of active users. To address these issues, we propose attention-enhanced prompt Decision Transformer (APDT).

As illustrated in Fig.~\ref{fig_2}, we first collect an offline dataset \(D_n\) for each \(\text{Env }n\), where each \(\text{Env }n\) corresponds to a specific environment configuration with a fixed number of users. This dataset stores optimal trajectories \(\tau_n\) containing transitions \(\bigl(s(t), a^*(t), r(t), c(t)\bigr)\), thus encapsulating expert-level decision-making under different user populations and offering strong supervision for subsequent training. 

The APDT is trained by minimizing the mean squared error between expert actions \(a^{\text{true}}\) and predicted actions \(a^{\text{pred}}\),
\begin{equation}
\label{loss_function}
L = \frac{1}{|D|} \sum_{\tau \in D} \bigl(a^{\text{true}} - a^{\text{pred}}\bigr)^2, \quad \theta \leftarrow \theta - \eta \,\nabla_\theta L,
\end{equation}
where \(\eta\) denotes the learning rate and \(|D|\) represents the size of the aggregated dataset from all scenarios. Training on multiple population configurations makes APDT robust and prevents unstable or looping flights when the user set changes.

\begin{algorithm}[!ht]
\caption{The attention-enhanced prompt Decision Transformer (APDT) training algorithm}
\label{alg:APDT}
\renewcommand{\algorithmicrequire}{\textbf{Input:}}
\begin{algorithmic}[1]
\REQUIRE APDT model $M_\theta$ with parameters $\theta$, offline dataset $\mathcal{D}$, pretraining prompt dataset $P$, segment length $K$, online buffer size $O$.
\STATE // Offline pre-training phase
\WHILE{not converged}
\STATE Initialize $\tau=\varnothing$
\FOR{Each Env $n=1,2,\dots,N$}
\STATE Sample a batch of trajectories $\mathcal{B}_n$ from $\mathcal{D}_n $
\STATE Random sample K-length prompt $p_n$ from $P_n$
\STATE Concatenate $\tau_n=\{p_n, B_n\}$, $\tau=\tau \cup \tau_n$
\ENDFOR
\STATE Calculate predict action $a^\text{pred}=M_\theta(\tau), \forall \tau \in \mathcal{B}$
\STATE Update APDT model using (\ref{loss_function})
\ENDWHILE
\STATE // Online deployment phase
\WHILE{not converged}
\STATE Observe initial state $s(1)$, target return $R$, cost $C$, random sample K-length prompt $p$ from online buffer $P^\prime$, initialize online trajectory $\tau^\prime=\{p, R, C, s(1) \}$
\FOR{$t \gets 1$ to $T$}
\STATE Compute action $a(t) = M_\theta(\tau^\prime)$
\STATE Execute $a(t)$, observe next state $s(t+1)$, reward $r(t)$ and cost $c(t)$
\STATE Update $R=R-r(t)$, $C=C-c(t)$, $\tau^\prime = \tau^\prime \cup (a(t), R, C, s(t+1))$ 
\ENDFOR
\STATE $P^\prime=P^\prime \cup \tau^\prime$

\IF{$|P^\prime| > O$} \label{line:check_buffer}
    \STATE Remove the oldest segment from $P^\prime$
\ENDIF

\ENDWHILE
\end{algorithmic}
\end{algorithm}

To achieve high returns while satisfying long-term energy constraint, the designed token-assisted method reconfigures the reward \( r(t) \) as the target return \( R(t) = \sum_{t' = t}^{T} r(t') \), and the cost \( c(t) \) as the cost-to-go \( C(t) = \sum_{t' = t}^{T} c(t') \). We then rearrange the trajectory sequence as $\tau(t)=(R(1), C(1), s(1), a^*(1), \dots, R(t), C(t), s(t), a^*(t))$. By this approach, we pre-train the APDT model on a dataset that satisfies long-term energy consumption constraint, enabling it to learn the mapping between future expected returns, long-term energy constraint, and action sequences.

To handle the varying dimensions of the state space arising from user arrivals and departures, we introduce an attention mechanism that treats the UAV’s position \(\boldsymbol{q}(t)\) as the “query” while each user’s position and Age of Information (AoI), \(\boldsymbol{s}_u(t) = \{q_u(t), A_u(t)\}\), serve as both “keys” and “values.” Formally, the attention operation is given by
\begin{equation}
\begin{aligned}[t]
&\text{Attention}\bigl(\boldsymbol{q}(t), \boldsymbol{s}_u(t), \boldsymbol{s}_u(t)\bigr) \\
&=\text{softmax}\!\Bigl(\frac{\bigl(\boldsymbol{W}_Q\,\boldsymbol{q}(t)\,\boldsymbol{W}_K\,\boldsymbol{s}_u(t)\bigr)^T}{\sqrt{d_k}}\Bigr)
\,\boldsymbol{W}_V\,\boldsymbol{s}_u(t),
\end{aligned}
\end{equation}
where $\mathbf W_Q\!\in\!\mathbb R^{d_k\times2}$ and
$\mathbf W_K,\mathbf W_V\!\in\!\mathbb R^{d_k\times3}$ project the 2‑D UAV
position and 3‑D user states to the same $d_k$‑dimensional space. First, APDT effectively handles dynamic user numbers through its attention mechanism, thereby ensuring decision stability and mitigating the potential decision biases of DT, which typically requires fixed-size inputs (e.g., via zero-padding). Second, in terms of training stability, APDT, similar to DT, employs a supervised learning based on offline data, which is inherently more stable than conventional DRL approaches, particularly when attention mechanisms are integrated online. Conventional DRL often suffers from increased complexity and convergence issue due to its reliance on real-time interaction and online joint training. In contrast, APDT benefits from offline pretraining, which eliminates the need for real-time feedback, thereby avoiding the challenges associated with online training and ensuring greater training stability.

We then incorporate a prompt-based mechanism for rapid adaptation, where a short demonstration segment of length \(K\), \(\bigl(\hat{R}(t), \hat{C}(t), \hat{s}(t), \hat{a}^(t)\bigr)\), is prepended to the input sequence. By providing APDT with partial‑trajectory context with environmental context information (e.g., the current user distribution), the prompt mechanism enables rapid adaptation to new scenarios without the need to retrain the model parameters $\theta$. This is a significant advantage in terms of efficiency, particularly in the dynamic UAV-AoI context, where traditional fine-tuning is often resource-intensive. This approach is not only computationally efficient \(K \ll T\)\, but also enhances generalization stability across diverse scenarios. Furthermore, compared to the implicit fine-tuning process of DT, the use of an explicit prompt provides a transparent basis for model adaptation. This explicit cue also clarifies the model’s adaptation logic and contributes to improved interpretability.

Algorithm \ref{alg:APDT} summarizes the APDT training process. In the offline pre-training phase, the model learns prior knowledge from expert dataset through supervised learning. During the online deployment phase, we initialize target return $R(1)$ and cost $C(1)$. The pre-trained model then interacts with the environment in real-time. Guidance prompts are randomly sampled from an experience buffer which is continually updated with recent $K$-length segments derived from interaction. This facilitates iterative adaptation and progressively improves decision-making performance until convergence.

\section{Simulation Results}
Simulations consider a UAV serving uplink users in an urban area. User positions are initially randomized and follow a Gauss–Markov mobility model \cite{liuPathPlanningUAVMounted2020}, though other mobility patterns are also compatible. The service area spans $500 \text{ m} \times 500 \text{ m}$ and the simulation runs for $T = 100$ time slots, each of duration $\delta = 5$ seconds. The UAV starts at position $q(1) = (100 \text{ m}, 100 \text{ m})$ with altitude $H = 60 \text{ m}$. Its per-slot flight distance is limited to $d^{\text{max}} = 90 \text{ m}$. The power consumption is $P^f=110 $ W during flight and $P^h=80W$ while hovering. The UAV's energy budget is capped at $E^{\max} = 90 \text{ KJ}$. Each user uploads a task of size $N_u = 30$ Mb with a power of $P_u = 0.5$ W. The channel bandwidth is $B = 1$ MHz, the power spectral density of Gaussian noise is $\sigma^2 = -140$ dB. The channel gain $h_u(t)$ is modeled as in~\cite{baiOpponentModelingBased2023}.


\begin{figure}[!t]
	\centering
	\includegraphics[width=2.5in]{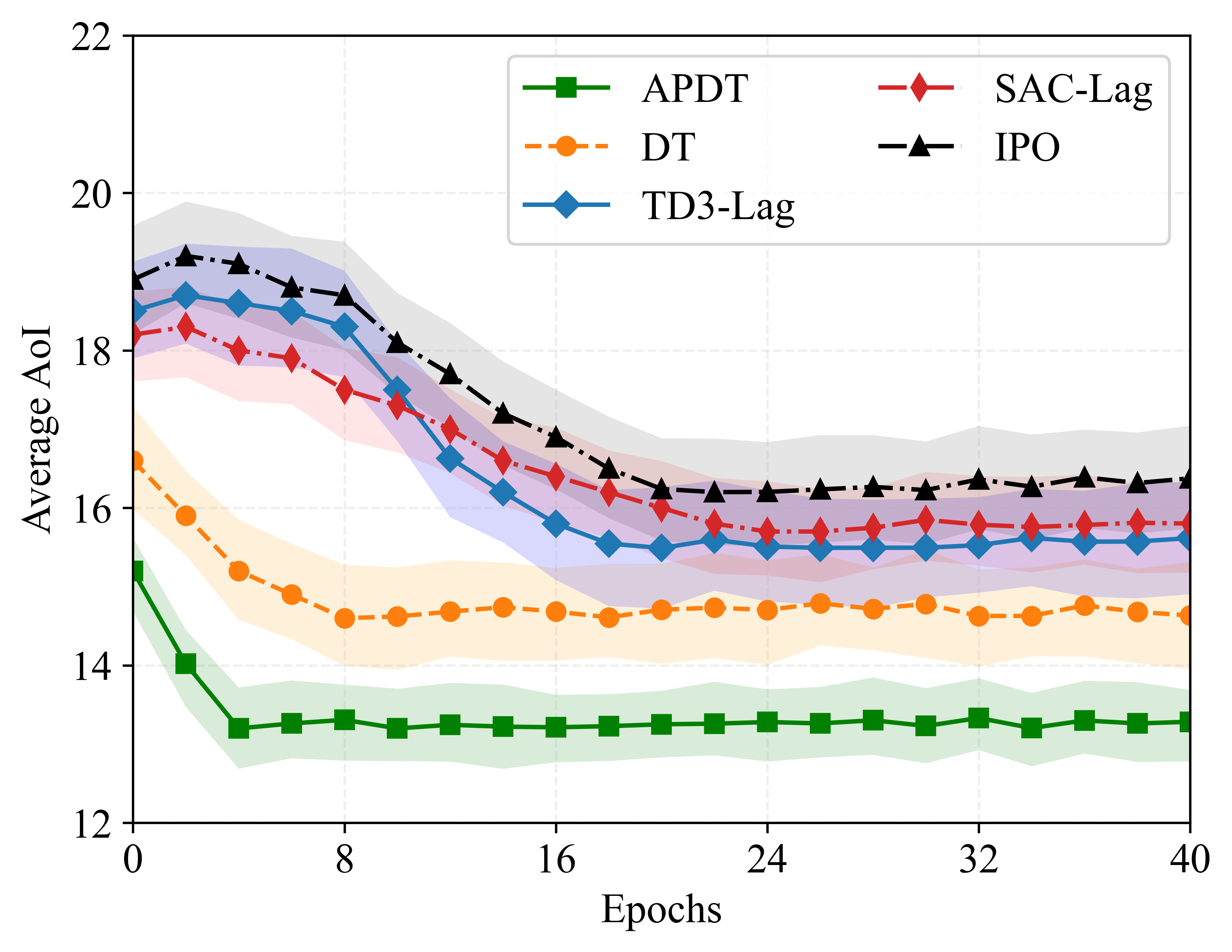}
	\caption{Average AoI performance of different algorithms in the new scenario, with energy constraint satisfied after convergence, except for DT.}
	\label{fig_3}
\end{figure}

\begin{figure}[!t]
	\centering
	\includegraphics[width=2.5in]{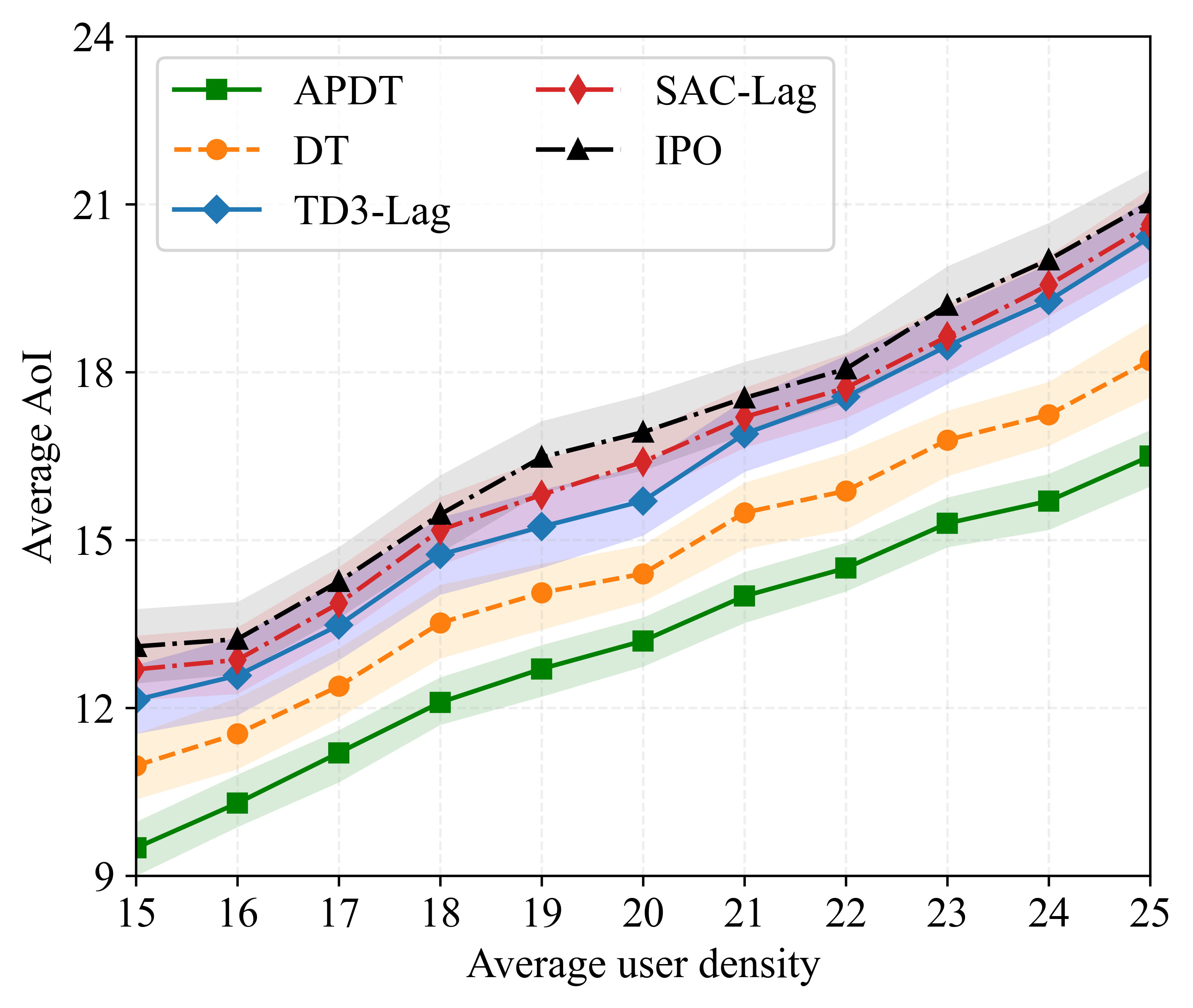}
	\caption{Average AoI performance of different algorithms across different average user densities.}
	\label{fig_4}
\end{figure}

Fig.~\ref{fig_3} compares the performance of APDT, DT, and conventional DRL algorithms, including Soft Actor-Critic with Lagrangian (SAC-Lag)~\cite{Achiam2019BenchmarkingSE}, Twin Delayed Deep Deterministic Policy Gradient with Lagrangian (TD3-Lag)\cite{Achiam2019BenchmarkingSE}, and Interior Point Optimization (IPO)~\cite{liuIPO2020} under varying user numbers with density $\rho=20$. Except for APDT, all algorithms pad the state space with zeros to match the state dimensionality. The pretraining datasets for both APDT and DT are generated from scenarios involving 11, 13, and 15 fixed users, with each dataset comprising 100,000 steps collected from a converged TD3-Lag agent. This data serves as the `expert' dataset, representing the optimal policy to which TD3-Lag effectively converges within these specific fixed-user environments. APDT uses 5‑step prompts and a $O=500$ online buffer, while DT performs real-time fine-tuning using samples generated by the TD3-Lag algorithm. For fairness, other algorithms were pre-trained in these scenarios before evaluation. All methods satisfy the energy threshold after convergence—except DT, which has no constraint mechanism.

The experimental results demonstrate that conventional DRL algorithms struggle in dynamic user scenarios, primarily due to their inability to handle varying state-space dimensions effectively and their lack of generalization capabilities to new environments. In contrast, APDT outperforms TD3-Lag from initialization, demonstrating effective knowledge acquisition from the offline dataset. Compared to conventional DRL, APDT achieves $4.5$ faster convergence and reduces average AoI by over $15\%$, highlighting its superior generalization capability and sample efficiency. This enhancement is primarily attributed to the prompt mechanism. Moreover, APDT demonstrates approximately $8\%$ lower average AoI and $2$ faster convergence compared to DT, validating the effectiveness of the attention and prompt mechanism.

To further evaluate the performance of the proposed method under varying user densities (with the average user density $\rho$ ranging from $15$ to $25$), we compared the AoI performance of different algorithms across various user densities, as shown in Fig.~\ref{fig_4}. As the number of users increases, the complexity of the environment grows, making it increasingly challenging for conventional DRL. This leads to a widening performance gap between the proposed method and conventional DRL. By leveraging offline dataset pretraining and direct deployment in new environments, the proposed method effectively overcomes these limitations. Additionally, the performance gap between APDT and DT also expands with the growing number of users, further validating the advantages of the attention mechanism in handling dynamic and complex environments.

\section{Conclusion}
We proposed an APDT framework for UAV-assisted communications, optimizing path planning and user scheduling to minimize average AoI while meeting energy constraint. By enhancing the DT framework with attention, prompt mechanisms, and a token-based approach, APDT demonstrates superior generalization ability to new scenarios with varying state-space dimensions, while satisfying long-term energy constraints. Simulations show that APDT achieves faster convergence and lower AoI than baseline methods. Future work will extend the framework to multi-UAV cooperative strategies.

\bibliographystyle{IEEEtran}
\bibliography{ref}

\end{document}